\documentclass[12pt]{article}

\usepackage{amssymb,amsmath}

\usepackage{graphicx}
\DeclareGraphicsExtensions{.pdf,.png,.jpg}

\renewcommand{\theequation}{\arabic{section}.\arabic{equation}}
 \catcode`\@=11 \@addtoreset{equation}{section}\catcode`\@=12
\newcommand{\R}{ {\mathbb R} }

\begin{document}

 \begin{center}

 \large \bf 
 Exponential cosmological solutions with two  factor spaces in EGB  model  with $\Lambda = 0$ revisited
  
  \end{center}

 \vspace{0.3truecm}

\begin{center}

 V. D. Ivashchuk$^{1,2}$ and A. A. Kobtsev$^{3}$ 

\vspace{0.3truecm}

 \it $^{1}$ 
   Institute of Gravitation and Cosmology, \\
   Peoples' Friendship University of Russia (RUDN University), \\
   6 Miklukho-Maklaya Street,  Moscow, 117198, Russian Federation, \\ 

 \it $^{2}$ Center for Gravitation and Fundamental Metrology,  VNIIMS, \\
 46 Ozyornaya Street, Moscow, 119361,  Russian Federation, \\
 
 \it $^{3}$ Institute for Nuclear Research of the Russian Academy of Sciences,
            Moscow, Troitsk, 142190,  Russian Federation

\end{center}

\begin{abstract}

We study exact  cosmological solutions  in   $D$-dimensional Einstein-Gauss-Bonnet model (with zero cosmological term) governed by  two non-zero constants:  $\alpha_1$ and  $\alpha_2$ .   
We deal with  exponential dependence (in time) of two scale factors governed by  Hubble-like parameters $H >0$ and $h$, which correspond to factor spaces of dimensions  $m >2$ and $l > 2$, respectively, and $D = 1 + m + l$. We put  $h \neq H$ and   $mH + l h \neq 0$. 
We show that  for  $\alpha  = \alpha_2/\alpha_1 > 0$ there are two (real) solutions with 
two sets of  Hubble-like parameters:  $(H_1, h_1)$ and  $(H_2, h_2)$, which obey:  
$ h_1/ H_1 < - m/l <  h_2/ H_2  < 0$, while for $\alpha < 0$ the (real) solutions are absent.   
We prove  that the cosmological solution corresponding to $(H_2, h_2)$    is stable  in 
a class of cosmological solutions with diagonal  metrics, while the  solution 
corresponding to  $(H_1, h_1)$ is unstable.  We present  several examples of analytical solutions, e.g.
stable ones with small  enough variation of the effective gravitational constant $G$, for $(m, l) = (9, l >2), (12,  11), (11,16), (15, 6)$.

\end{abstract}


\section{Introduction}

Currently, the Einstein-Gauss-Bonnet (EGB)  model and  related theories, see \cite{Ishihara}-\cite{IvKob-18-3l}
and refs. therein,  are under intensive studies in  cosmology,
aimed at   explanation  of  accelerating  expansion of the Universe  \cite{Riess,Perl}.
Here  we  study the EGB model with zero cosmological term  in $D$ dimensions ($D = n+1$).  This model contains Gauss-Bonnet term, which  arises  in (super)string theory as a correction to the  (super)string effective action (e.g. heterotic one) \cite{Zwiebach}-\cite{GW}. The model is governed by two nonzero constants
$\alpha_1$ and $\alpha_2$ which correspond to Einstein and Gauss-Bonnet  terms in the action, respectively.
In this paper we continue our studies of  the EGB cosmological model  from ref. \cite{IvKob}. We deal with diagonal metrics  governed by $n >3$ scale factors and  consider the following ansatz for scale factors
$a_i(t)$ ($t$ is synchronous time variable): $ a_1(t) = \dots = a_m(t) = \exp(Ht)$ and 
$a_{m+1}(t) = \dots = a_{m+l}(t) = \exp(ht)$, where $n =m + l$,  $m > 2$, $l > 2$. We put here $H >0$ in order
to describe exponential accelerated expansion  of $3d$ subspace with Hubble parameter $H$ \cite{Ade}.

In contrary to our earlier publication \cite{IvKob}, where a lot of numerical solutions
with small enough value of variation of the effective  gravitational constant $G$ were found, 
here we put our attention mainly to the search of analytical exponential solutions with two factor spaces of dimensions $m$ and $l$.
Here we show that the anisotropic cosmological solutions
under consideration  with two Hubble-like parameters $H>0$ and $h$
obeying  restrictions  $h \neq H$,  $mH + l h \neq 0$  do exist only if $\alpha = \alpha_2/ \alpha_1> 0$. In this case we have
two solutions with Hubble-like parameters: $(H_1 > 0, h_1<0)$ and $(H_2 > 0, h_2<0)$, respectively, such that
$x_1 = h_1/H_1<  - m/l  <  x_2 = h_2/H_2$. 
By using results of  refs. \cite{ErIvKob-16,Ivas-16} 
(see also approach of ref. \cite{Pavl-15})  we show  that the solutions 
with Hubble-like parameters $(H_2, h_2)$ are stable (in a class of cosmological solutions 
with diagonal metrics), while those corresponding to $(H_1, h_1)$ are unstable.

Here we also present  examples of analytical solutions for: i)  $m =l$;  ii) $m=3$, $l =4$;
iii) $m = 9$, $l >2$; iv) $m = 12$, $l = 11$; v) $m=11$, $l=16$ and vi) $m =15$, $l =6$.
It should be noted that analytical solutions in cases iii) and   iv) were considered numerically 
in ref. \cite{IvKob} in a context of solutions with a small (enough)  variation  of  $G$  (in Jordan  frame, see ref. \cite{RZ-98}), e.g. obeying the most severe restrictions on  variation of $G$ from ref. \cite{Pitjeva}.
The stable solutions with zero variation  of  $G$  in cases v)  and vi) were found earlier in \cite{IvKob},
while the stability of these solutions was proved in ref. \cite{ErIvKob-16}.

\section{The set up}

We start with the following action of the model 
\begin{equation}
  S =  \int_{M} d^{D}z \sqrt{|g|} \{ \alpha_1 R[g] +
              \alpha_2 {\cal L}_2[g] \}.
 \label{r1}
\end{equation}
Here $g = g_{MN} dz^{M} \otimes dz^{N}$ is the metric defined on
the manifold $M$, ${\dim M} = D$, $|g| = |\det (g_{MN})|$, $\Lambda$ is
the cosmological term, $R[g]$ is scalar curvature,
$${\cal L}_2[g] = R_{MNPQ} R^{MNPQ} - 4 R_{MN} R^{MN} +R^2$$
is the  Gauss-Bonnet term and  $\alpha_1$, $\alpha_2$ are
nonzero constants.

We deal with  warped product manifold
\begin{equation}
   M = \R  \times   M_1 \times \ldots \times M_n 
   \label{r2.1}
\end{equation}
with the (cosmological) metric
\begin{equation}
 g= - dt \otimes dt + \sum_{i=1}^{n} e^{2\beta^i(t)}  dy^i \otimes dy^i,
 \label{r4.3a}
\end{equation}
 where $M_1, \dots,  M_n$  are one-dimensional manifolds (either $\R$ or $S^1$)
and $n > 3$.

Here we put 
\begin{equation}
\beta^i(t)  = v^i t + \beta^i_0,
 \label{r4.3b}
\end{equation}
$i = 1, \dots, n$, where $v^i$ and $\beta^i_0$ are constants. 

The equations of motion for the action (\ref{r1}) 
give us the set of  polynomial equations \cite{Iv-09,Iv-10}
\begin{eqnarray}
  G_{ij} v^i v^j   - \alpha   G_{ijkl} v^i v^j v^k v^l = 0,  \label{r2.3} \\
    \left[ 2   G_{ij} v^j
    - \frac{4}{3} \alpha  G_{ijkl}  v^j v^k v^l \right] \sum_{k =1}^n v^k 
    - \frac{2}{3}   G_{sj} v^s v^j = 0,
   \label{r2.4}
\end{eqnarray}
$i = 1,\ldots, n$, where  $\alpha = \alpha_2/\alpha_1$. 
Here we denote \cite{Iv-09,Iv-10}
\begin{equation}
G_{ij} = \delta_{ij} -1, 
\qquad   G_{ijkl}  = G_{ij} G_{ik} G_{il} G_{jk} G_{jl} G_{kl}.
\label{r2.4G}
\end{equation}
For the case $n > 3$  (or $D > 4$) we have a set of forth-order polynomial  equations.


\section{Solutions governed by two Hubble-like parameters}

Here we study solutions to  equations (\ref{r2.3}), 
(\ref{r2.4}) with following set of Hubble-like parameters
\begin{equation}
  \label{r3.1}
   v =(\underbrace{H,H,H}_{``our" \ space},\underbrace{\overbrace{H, \ldots, H}^{m-3}, 
   \overbrace{h, \ldots, h}^{l}}_{internal \ space}).
\end{equation}
where $H$ is the Hubble-like parameter corresponding  
to an $m$-dimensional factor space with $m > 2$, while  $h$ is the Hubble-like parameter 
corresponding to an $l$-dimensional factor space, $l > 2$. 
The splitting in  (\ref{r3.1}) was done just  for  
cosmological applications. Here we split the $m$-dimensional  
factor space into the  product of $3d$ subspace (``our"  space) and $(m-3)$-dimensional subspace, which
is a part of $(m-3 +l)$-dimensional ``internal'' space.
 
Keeping in mind a possible description of an accelerated expansion of a
$3d$ subspace,  we impose the following restriction 
\begin{equation}
  \label{r3.2a}
   H > 0. 
\end{equation}
  
Due to  ansatz (\ref{r3.1}),  the $m$-dimensional subspace is expanding with the Hubble parameter $H >0$. 
The  behaviour of scale factor corresponding to $l$-dimensional subspace 
is governed  by  Hubble-like  parameter $h$.

Here we use the results of refs. \cite{ChPavTop1,Ivas-16} which 
 tell us that the imposing of  two  restrictions on  $H$ and $h$  
   \begin{equation}
   m H + lh \neq 0, \qquad  H \neq h,
   \label{r3.3}
   \end{equation}
  reduces  (\ref{r2.3}) and (\ref{r2.4})  to the  set 
  of two (polynomial) equations
  \begin{eqnarray}
  E =  m H^2 + l h^2 - (mH + lh)^2  
        - \alpha [m (m-1) (m-2) (m - 3) H^4
          \nonumber \\
       + 4 m (m-1) (m-2) l H^3 h   
       + 6 m (m-1) l (l - 1) H^2 h^2
         \nonumber \\
       + 4 m l (l - 1) (l - 2) H h^3  
       + l (l - 1) (l - 2) (l - 3) h^4] = 0, \quad
         \label{r3.4}   \\
  Q =  (m - 1)(m - 2)H^2 + 2 (m - 1)(l - 1) H h 
        \nonumber \\ 
       + (l - 1)(l - 2)h^2 = - \frac{1}{2 \alpha}.
     \label{r3.5}
  \end{eqnarray}

Relation (\ref{r3.5}) implies for $m > 2$ and $l > 2$: 
\begin{equation}
H   =     (- 2 \alpha {\cal P})^{-1/2}, 
 \label{r3.6}
\end{equation}
where 
\begin{eqnarray}
{\cal P}  = {\cal P}(x,m,l) \equiv  (m - 1)(m - 2) 
  \nonumber \\  
  + 2 (m - 1)(l - 1) x  + (l - 1)(l - 2)x^2, 
 \label{r3.7}  \\
    x  = h/H,
    \label{r3.7x}
 \end{eqnarray}
and 
 \begin{equation}
\alpha {\cal P} < 0. 
 \label{r3.8} 
\end{equation}

We rewrite  (\ref{r3.3})  as follows
\begin{equation}
  x \neq x_d  \equiv  - m/l, \qquad  x \neq x_a \equiv 1. 
 \label{r3.8da} 
\end{equation}

  The relation  (\ref{r3.8}) lead us to inequality        
\begin{equation}
  {\cal P}(x,m,l) \neq 0. 
 \label{r3.8b}
\end{equation}

Using (\ref{r3.4}) and (\ref{r3.6})
we obtain
\begin{eqnarray}
   \lambda(x)  =  \lambda(x,m,l) \equiv 
     \frac{1}{4} ({\cal P}(x,m,l))^{-1} {\cal M}(x,m,l)
     \qquad \nonumber \\
      + \frac{1}{8 }( {\cal P}(x,m,l))^{-2} {\cal R}(x,m,l) = 0,  
              \qquad    \label{r3.8L}  \\
   {\cal M}(x,m,l) \equiv  m  + l x^2 -(m  + l x)^2, 
              \qquad   \label{r3.8M}  \\
   {\cal R}(x,m,l) \equiv  m (m-1) (m-2) (m - 3)
                 + 4 m (m-1) (m-2) l x 
                \qquad  \nonumber \\    
       + 6 m (m-1) l (l - 1) x^2        
       + 4 m l (l - 1) (l - 2)  x^3
        \qquad  \nonumber \\
       + l (l - 1) (l - 2)(l - 3) x^4 = 0.
               \qquad   \label{r3.8R}  
\end{eqnarray}

 Here the following identity is valid 
 \begin{equation}   
    \lambda(x,m,l) = \lambda (1/x,l,m) 
               \qquad   \label{r3.9l}   
 \end{equation}
 for $x \neq 0$.

It follows from (\ref{r3.8b}) that \cite{IvKob-18ml}
 \begin{eqnarray} 
  x \neq x_{\pm}  
  \equiv \frac{-(m - 1)(l - 1) \pm \sqrt{\Delta}}{(l - 1)(l - 2)}, 
 \label{r3.9} \qquad \\
   \Delta \equiv  (m - 1)(l - 1)(m + l - 3),
            \qquad \label{r3.9D}
 \end{eqnarray}
where $x_{\pm}(m,l)$ are roots of the quadratic equation ${\cal P}(x,m,l) =0$,
 obeying 
\begin{equation} 
    x_{-} < x_{+} < 0. 
         \qquad \label{r3.11} 
  \end{equation}

Using (\ref{r3.8}) we get
\begin{equation} 
    x_{-} < x <  x_{+} \ \ {\rm for} \  \alpha > 0,  
               \qquad \label{r3.13b}
 \end{equation}    
  and    
   \begin{equation}  
     x <  x_{-} \ {\rm or} \ x > x_{+} \ \ {\rm for} \  \alpha < 0.
                 \qquad \label{r3.13c}
 \end{equation}

For $ \alpha < 0$ the following relation is valid 
  \begin{equation}  
  \lim_{x \to \pm \infty} \lambda(x,m,l) = 
   \lambda_{\infty}(l) \equiv  - \frac{l(l + 1)}{8 (l - 1)(l - 2)} < 0.
                  \qquad \label{r3.13.l}
  \end{equation}

 Equation (\ref{r3.8L}) may be rewritten in the following form
\begin{equation}
       2 {\cal P}(x,m,l) {\cal M}(x,m,l) +  {\cal R}(x,m,l)  = 0,  
              \qquad    \label{r3.13.M} 
\end{equation}                
or, equivalently,
\begin{eqnarray}
  l(l-1)(l-2)(l-3)x^4 + 4ml(l-1)(l-2)x^3 
 \nonumber \\   
+ 6 m(m-1)l(l-1)x^2 + 4m(m-1)(m-2)lx 
\nonumber \\
+ m(m-1)(m-2)(m-3) 
 \nonumber \\
  + 2[(m-1)(m-2)+ 2(m-1)(l-1)x 
 \nonumber \\
 +(l-1)(l-2)x^2 ][m+lx^2-(m+lx)^2] =0. 
\label{r3.13.Ma} 
\end{eqnarray}

This equation  is of fourth order in $x$ for any $l > 2$. 
One can solve the  equation (\ref{r3.13.Ma}) in radicals for any $m > 2$ and $l > 2$. 
The general  solution is  presented in Appendix.

 Here we use the following proposition from ref. \cite{IvKob-18ml}.

{\bf Proposition 1 \cite{IvKob-18ml}.} {\em For $m > 2$, $l > 2$ 
 \begin{equation}
    \lambda(x,m,l) \sim B_{\pm} (x - x_{\pm})^{-2}, 
 \label{r3.13.H}
\end{equation}
 as $x \to x_{\pm}$, where $B_{\pm} < 0$ and hence
 \begin{equation}
   \lim_{x \to x_{\pm}}  \lambda(x,m,l) = - \infty. 
  \label{r3.13.lim}
 \end{equation}
 }

 In what follows we use the relations for the  extremum  points of the function 
 $\lambda(x)$  ($\frac{\partial}{\partial x} \lambda(x,m,l) = 0$) from
 \cite{IvKob-18ml}:
 \begin{eqnarray} 
        x_a = 1,
            \qquad \label{r3.15a} \\
        x_b =   - \frac{m-1}{l-2} < 0,
             \qquad \label{r3.15b} \\
        x_c =  - \frac{m-2}{l-1} < 0,
                     \qquad \label{r3.15c} \\
        x_d =  - \frac{m}{l} < 0,
                             \qquad \label{r3.15d}       
 \end{eqnarray}
which follow from the identity \cite{IvKob-18ml}
\begin{eqnarray} 
  \frac{\partial}{\partial x} \lambda(x,m,l) = - f(x,m,l) ({\cal P}(x,m,l))^{-3},
                \qquad \label{r3.14a} \\
        f(x,m,l) = (l-1)(m-1)(x-1)(lx+m) \times
                 \nonumber \\
        \times [(l-2)x + m-1][(l-1)x + m-2],
                 \qquad \label{r3.14b}
\end{eqnarray}
$x \neq x_{\pm}$.

Here $x_b < x_c$ and       
the points $ x_b, x_c, x_d$ belong to the interval 
   $(x_{-},x_{+})$ for all $m > 2$ and $l > 2$.
The location of the point $x_d$ depends upon
$m$ and $l$ \cite{IvKob-18ml}: 
\begin{eqnarray}
(1) \ x_b < x_c < x_d & \text{for } l < m/2,  \label{r3.20a} \\
(2) \ x_b < x_d < x_c & \text{for }  m/2 < l < 2m, \label{r3.20b} \\
(3) \ x_d < x_b < x_c & \text{for } l > 2m, \label{r3.20c}
\end{eqnarray}	
and 
\begin{eqnarray}
(1_0) \ x_b < x_c = x_d & \text{for } l = m/2, \label{r3.21a} \\
(3_0) \ x_d = x_b < x_c & \text{for } l = 2m. \label{r3.21b}
\end{eqnarray}

The values $\lambda_i=\lambda(x_i,m,l)$, $i=a,b,c,d$, were calculated 
in \cite{IvKob-18ml}. They obey  
\begin{equation}
  \lambda_{\infty} = \lambda_{\infty}(l) < \lambda_a <0, \qquad \lambda_i>0,          \label{r3.22a}
\end{equation}
$i = b, c,d$.

 First, we consider  the case $\alpha > 0$ and $x_{-} < x < x_{+}$. 
 
\begin{figure}[!h]
	\begin{center}
		\includegraphics[width=0.75\linewidth]{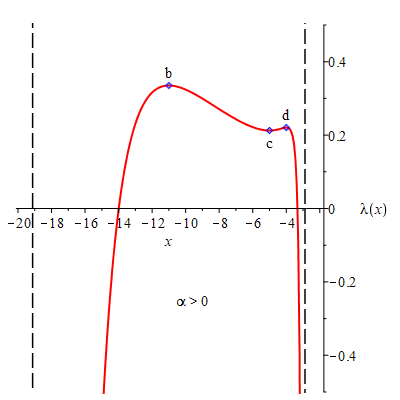}
		\caption{The function $\lambda(x)$ for $\alpha > 0$,  $m=12$ and $l=3$ \cite{IvKob-18ml}.}
		\label{rfig:1}
	\end{center}
\end{figure}

For $\alpha > 0$ in cases  $(1), (2)$ and $(3)$ we have two points of local maximum and one point of local minimum
among $x_b,  x_c$ and  $x_d$, see Figure 1,  while in cases $(1_0)$ and $(3_0)$ we have one point of local maximum and one point of inflection, see Figure 2. 
Due to relations (\ref{r3.14a}), (\ref{r3.14b}) the function $\lambda(x)$ is monotonically increasing in the interval 
$(x_{-}, {\rm min}(x_b, x_c, x_d))$, and it is  monotonically decreasing in the interval $({\rm max}(x_b, x_c, x_d), x_{+})$.

 \begin{figure}[!h]
  	\begin{center}
  		\includegraphics[width=0.75\linewidth]{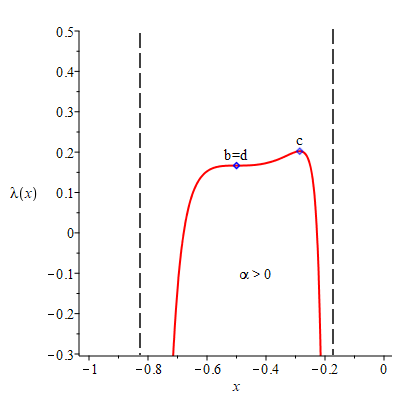}
  		\caption{The function $\lambda(x) $ for $\alpha > 0$, $m =4$ and $l = 8$ \cite{IvKob-18ml}.}
  		\label{rfig:2}
  	\end{center}
  \end{figure}

Now, let us consider the case  $\alpha < 0$. 
We have: $x < x_{-}$ or $x > x_{+}$.  Due to to 
the relations (\ref{r3.13.l}), (\ref{r3.14a})  and Proposition 1, 
the function $\lambda(x)$  is monotonically decreasing in two intervals: i) in the
interval $(-\infty,x_{-})$ from $ \lambda_{\infty} $ to $- \infty$ and ii) in the
interval $(x_{a} = 1, +\infty)$ from $ \lambda_{a} $ to $ \lambda_{\infty} $. 
The function $ \lambda(x)$  is monotonically increasing in the  interval  
$(x_{+}, x_{a})$ from $- \infty$ to $\lambda_{a} $.
Here $x_{a}= 1$ is a point of local maximum of the function $\lambda(x) $, which
is excluded from the solution and   $0 >  \lambda_{a} > \lambda_{\infty}$.  
 
 The  functions 
 $\lambda(x)/\alpha$   for $\alpha = +1, -1$, respectively, and $m = l =4$ are 
 presented at Figure 3.

 \begin{figure}[!h]
   	\begin{center}
   		\includegraphics[width=0.80\linewidth]{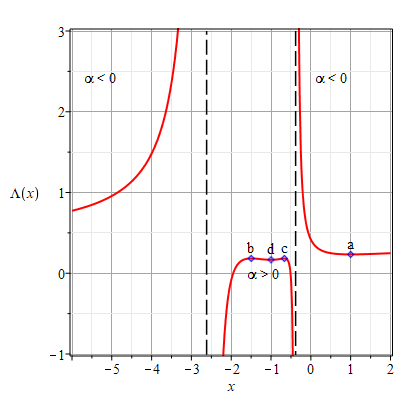}
   		\caption{The functions $\Lambda(x) = \lambda(x)/\alpha $ for $\alpha = \pm 1 $
   		and $m = l = 4$ \cite{IvKob-18ml}.}
   		\label{rfig:3}
   	\end{center}
   \end{figure}

By using the behaviour of the function $\lambda(x,m,l)$, which was considered above, 
one can readily prove the following proposition.

{\bf Proposition 2.} {\em For any $m > 2$, $l > 2$ there are only two 
real solutions $x_1, x_2 $ to the master equation $\lambda(x) = \lambda(x,m,l) = 0$ 
(see (\ref{r3.8L})) for $\alpha > 0$. These 
solutions obey $x_{-} < x_1 < - \frac{m}{l} < x_2 < x_{+} < 0$ (see  (\ref{r3.9})). For  
 $\alpha < 0$ the solutions to master equation are absent.  }

{\bf Proof.}
First, let us consider  the case  $\alpha < 0$.  In this case it follows from our analysis above that  $\lambda(x) < \lambda_{\infty}$ 
for $x < x_{-}$ and $\lambda(x) < \lambda_{a}$. Since $\lambda_{\infty} <  \lambda_a < 0$, we get in the case 
$\alpha < 0$:  $\lambda(x) < \lambda_{a} < 0$. Hence the equation $\lambda(x) = 0$ does not have solutions. 

Now we consider the case  $\alpha > 0$. We are seeking the solutions to equation $\lambda(x) = 0$ in the interval $(x_{-}, x_{+})$,
where our function is smooth (and continuous).
Let us denote: $x_{*} =  {\rm min}(x_b, x_c, x_d)$ and  $x_{**} ={\rm max}(x_b, x_c, x_d)$.
The interval  $[x_{*},x_{**}]$ should be excluded from our consideration 
since $\lambda(x) \geq {\rm min}(\lambda_a,\lambda_b,\lambda_c) > 0$  for $x \in [x_{*},x_{**}]$.
(Here we use the fact that the smooth (e.g. continuous) function on the closed interval $[x_{*},x_{**}]$ has a minimum which should
be equal to $\lambda(x_{*})$ or $\lambda(x_{**})$ or a value of the function
in a point of local minimum (e.g. point of extremum) of the form $\lambda(x_{i})$, $i = b,c, d$. 
In any case this minimum coincides with $\lambda(x_{i})$  for some $i = b,c, d$.)
Now we consider the interval $(x_{-},x_{*})$. The function $\lambda(x)$ is monotonically increasing in the interval  $(x_{-},x_{*})$. 
Due to relation (\ref{r3.13.l}) there exists a point $x_{*,-} \in (x_{-},x_{*})$
such that $\lambda(x_{*,-}) < -1$ and hence any point $x$ in the interval  $(x_{-}, x_{*,-}]$
obey $\lambda(x) < -1$. Thus, we exclude the interval $(x_{-}, x_{*,-}]$ from our consideration. Now we 
consider the interval $[x_{*,-}, x_{*}]$, where $\lambda(x_{*,-}) < -1$ and $\lambda(x_{*}) > 0$.
Due to intermediate value theorem there exists a point $x_1 \in (x_{*,-}, x_{*}) \subset (x_{-}, x_{*})$  such that 
$\lambda(x_{1}) = 0$. This point is unique since the function is monotonically increasing in this interval.
By analogous arguments one can readily prove the existence of unique point $x_2 \in (x_{*}, x_{+})$ such that 
 $\lambda(x_{2}) = 0$.  By our definitions above we obtain   
 $x_{-} <x_1 < x_{*} \leq x_d  = - \frac{m}{l} \leq  x_{**} < x_2 < x_{+} < 0$. This completes the proof of the proposition. 

Thus, we are led to the following (physical) result: the anisotropic cosmological solutions 
under consideration  with two Hubble-like parameters $H>0$ and $h$
obeying  restrictions  (\ref{r3.3}) do exist only if $\alpha > 0$. In this case we have 
two solutions with Hubble-like parameters: $(H_1 > 0, h_1 <0)$ and $(H_2 > 0, h_2 <0)$ such that 
$h_1/H_1  <  - m/l  <  h_2/H_2< 0$.

\section{Stability analysis and variation of $G$ }

 Now, we consider the stability of cosmological solutions in a class of solutions 
 with the metric (\ref{r4.3a})
\begin{equation}
 g= - dt \otimes dt + \sum_{i=1}^{n} e^{2\beta^i(t)}  dy^i \otimes dy^i.
 \label{r4.3}
\end{equation}

In ref. \cite{IvKob-18ml}  we have proved the following proposition, which is valid 
for exponential solutions with two factor spaces and Hubble-like parameters obeying  
(\ref{r3.2a}) and (\ref{r3.3})  in the EGB model with a $\Lambda$-term:  

{\bf Proposition 3 \cite{IvKob-18ml}.} 
{\em The cosmological solutions  from \cite{IvKob-18ml}, which obey  $x = h/H \neq x_i$, 
$i = a,b,c,d$, where $x_a =1$, $x_b = - \frac{m -1}{l - 2}$, $x_c = - \frac{m-2}{l -1}$,  $x_d = - \frac{m}{l}$, 
are  stable, if i) $x > x_d$ and unstable, if ii) $x < x_d$. }

Here it should be noted that our anisotropic solutions with non-static volume factor
are not defined for $x= x_a$ and $x= x_d$. Meanwhile, they
are defined when $x = x_b$ or $x = x_c$, if $x \neq x_d$. 

{\bf Proposition 4.} {\em The cosmological solution under consideration for $\alpha >0 $ corresponding 
to the big root of master equation $x_2 $ is  stable,  while the solution corresponding 
to the small root $x_1$ is unstable. }

 Here we analyze the solutions  by using the restriction on variation of the effective
 gravitational constant $G$ (in the Jordan frame), which is inversely proportional 
 to the  volume scale factor  of the (anisotropic) internal space  
 (see \cite{IvKob} and references therein), i.e.

 \begin{equation}
  \label{r5.G0}
    G = {\rm const } \exp{[- (m - 3) H t - l h t]}.
 \end{equation}
 
By using (\ref{r5.G0}) we get 
\begin{equation}
  \delta \equiv \frac{\dot{G}}{GH}  =  - (m - 3 + l x), \qquad x = h/H.
\label{r5.G}
\end{equation}

Here we use, as in ref. \cite{IvKob},  the following bounds on the 
value of the dimensionless variation of the  effective gravitational constant:

 \begin{equation}
 \label{r5.G1}
  - 0.65 \cdot 10^{-3} < \frac{\dot{G}}{GH} < 1.12 \cdot 10^{-3}.
 \end{equation}
They come from the most stringent bounds
on $G$-dot (by the set of ephemerides) \cite{Pitjeva}
 $\dot{G}/G = (0.16 \pm 0.6) \cdot 10^{-13} \ year^{-1}$,
which are allowed at 95\% confidence (2-$\sigma$) level,
and the  value of the Hubble parameter (at present) \cite{Ade}
  $H_0 =  (67,80 \pm 1,54) \ km/s \ Mpc^{-1} =  (6.929  \pm 0,157) \cdot 10^{-11} \ year^{-1}$,
  with 95\% confidence level. 
  
   Let us consider the solution with $x$-parameter corresponding to
  dimensionless parameter  of variation of $G$  from (\ref{r5.G}).
  Then, we have 
    \begin{equation}
  x =  x_0(\delta,m,l) \equiv  - \frac{(m - 3 + \delta)}{l} 
  \label{r5.0}    
 \end{equation}
 and 
  \begin{equation}
  x_0(\delta,m,l) - x_d =  \frac{  3 -  \delta}{l} > 0 
  \label{r6.1}    
 \end{equation} 
 for 
 \begin{equation}
   \delta < 3.   \label{r6.2}    
  \end{equation}
 
 Let us consider a solution  with a small enough parameter $\delta$, which satisfies  
 restrictions (\ref{r5.G1}). It obeys  (\ref{r6.2}) and hence we obtain from 
 (\ref{r6.1})  $x = x_2$ 
 since $ x_d < x_2$, while  $x_1 < x_d$.  Thus, this solution is stable due to 
 Proposition 4. Hence, all solutions with small enough variation of $G$,
 which were obtained in ref. \cite{IvKob}, are stable. 
 The stability two of them was proved in ref. \cite{ErIvKob-16}.
 
{\bf Remark.} It follows from our consideration that a more wide class of solutions with $\delta < 3$ consists of stable solutions.

   \section{Examples of solutions}
   
   Here we present certain examples of analytical solutions in the model under consideration.
   These solutions may be readily verified by using Maple or Mathematica. 
   They  are given by $x = x_1, x_2$ and relations (\ref{r3.6}), (\ref{r3.7}), (\ref{r3.7x}).

   \subsection{The  solutions for $m=l$}
   
   For any $m = l > 2$ the master equation (\ref{r3.13.M}) was solved in fact in ref. \cite{IvKob-18mm} 
   (it was solved there for arbitrary $\Lambda$). The solution reads 
    
    \begin{eqnarray}
    \begin{aligned}
       x(\nu,m) = ((m+1)(m-2))^{-1} \left( -   (m-1)^2  - \sqrt{2m^2-7m+7} \right.  \nonumber \\  
      \left. 
     + \nu   \sqrt{ - (2 m^3- 11 m^2+15 m-4) + 2 (m-1)^2 \sqrt{2m^2-7m+7}}  \right),    
   \label{A.1}
   \end{aligned}
   \end{eqnarray}
   $m > 2$, $\nu = \pm 1$. In our notations  $x_1 = x(-1,m)$ and $x_2 = x(1,m)$.
   
   For $m = 3,4,5$ we get:
   \begin{equation}
    x(\nu,3) = \frac{1}{2} (- 3 + \nu  \sqrt{5}),
         \label{A.2} 
    \end{equation} 
    see \cite{IvKob-18-3l}, and 
    \begin{equation}
     x(\nu,4) = \frac{1}{10} \left( - 9 - \sqrt{11} 
      + \nu   \sqrt{ 18 \sqrt{11} - 8}  \right),    
     \label{A.3}
    \end{equation}
    
   \begin{equation}
     x(\nu,5) = \frac{1}{18} \left( - 16 - \sqrt{22} 
      + \nu   \sqrt{ 32 \sqrt{22} - 46}  \right).    
     \label{A.4}
    \end{equation}

   \subsection{The solution for $m=3$ and  $l =4$}
   
      For the case $m= 3$, $l =4$ the master equation (\ref{r3.13.M}) has two real solutions
         \begin{eqnarray}
   x(\nu) =  - \frac{1}{30}  X^{-1/6} Y^{1/2} -3/5   \nonumber \\
    + \frac{\nu}{2} \sqrt{ \frac{216}{25} X^{1/6} Y^{-1/2} - X^{1/3}+ \frac{7}{45} ( X^{-1/3}- 1)},
        \label{B.1}   \\     
            X = \frac{14 \sqrt{13}}{375 \sqrt{3}} + \frac{161}{3375},     \label{B.2}   \\                
            Y = 225 X^{2/3}-6 X^{1/3}-35,  \label{B.3}
        \end{eqnarray}
   $\nu = \pm 1$. In our notations $x_1 = x(-1)$ and $x_2 = x(1)$. 
   (Approximate values are following ones: $x_1 = -1,345775$ and $x_2 = - 0,258116$.)
   
 \subsection{The series of solutions for $m=9$ and $l > 2$}

Now we consider the case $m=9$, $l > 2$.
The master equation (\ref{r3.13.M}) in this case  reads
\begin{eqnarray}
(l-2)(l-1)l(l+1) x^4 + 32(l-1)^2 lx^3  \nonumber \\
+16(l-1)(25l-18)x^2 + 2304(l - 1)x + 5040 = 0.
\label{C.0}
 \end{eqnarray} 
 It has two real solutions for any $l > 2$
  \begin{eqnarray}
     x(\nu,l) =  - \frac{1}{M} X^{-1/6} Y^{1/2} - R  + \frac{\nu}{2} Z^{1/2},
          \label{C.0}   \\ 
    Z =  P_Z X^{1/6} Y^{-1/2} - X^{1/3}+ Q_Z X^{-1/3}- R_Z,     \label{C.1}   \\
    Y = N_0 X^{2/3}- N_1 X^{1/3}- N_2,  \label{C.2} \\
    X = P_X \sqrt{Q_X} + R_X,     \label{C.3}  
   \end{eqnarray}  
 where $\nu = \pm 1$ and
   \begin{eqnarray}                           
 N_0 =  9(l-2)^2 (l-1)l^2(l+1)^2, \label{C.N0}   \\
 N_1 =  96(l-1)l(l^3+5l^2-56l+36),  \label{C.N1} \\                  
 N_2 =   64(l+9)(11l^2+34l+144),  \label{C.N2} \\ 
 R = \frac{8(l-1)}{(l-2)(l+1)},        \label{C.R} \\                
 M = 6l(l-2)(l+1) \sqrt{l-1},     \label{C.M} \\                                   
 P_Z  = \frac{3072(l-1)^{1/2}(l^3+5 l^2-24l+36)}{(l-2)^2(l+1)^2}, \label{C.PZ} \\
 Q_Z = \frac{N_2}{N_0}, \qquad    \qquad   \qquad    \qquad       \label{C.QZ} \\
 R_Z =  \frac{64 (l^3+5l^2-56l+36)}{3 l (l^2 - l - 2)^2}   \label{C.QZ} \\
 P_X = \frac{512(l+9)}{(l-1)l^2(l-2)^3(l+1)^3}       \label{C.PX} \\ 
 Q_X =  \frac{(l+6)(l+8)(5 l^2+4l+36)(9l^2+17l+72)}{3l(l-1)},                  \label{C.QX} \\  
 R_X = \frac{1024 (l+9)(49l^3+428l^2+900l+2592)}{27(l-2)^3(l-1)l^3 (l+1)^3} \label{C.RX}. \\ 
\end{eqnarray} 

Now, we study the behaviour of solutions $x_1 = x(-1,l)$ and $x_2 = x(1,l)$ for big values of $l$.
By using $(1/l)$-decomposition we get 
 \begin{eqnarray}
 x_1 = - \frac{10}{l} + o(l^{-1}),  \label{C.4} \\
 x_2 = - \frac{6}{l} + o(l^{-1}),
 \label{C.5}
 \end{eqnarray}
 for $l \to \infty$.
 These relations just follow from the formulae
  \begin{eqnarray}
   X  = X_{\infty} l^{-6} ( 1 + o(l^{-1})),   \label{C.6} \\
   \bar{Y} \equiv Y X^{-1/3}  = 2^{12} l^{3} (1 + o(l^{-1}))
   \label{C.7}
  \end{eqnarray}
   as $l \to \infty$, where 
   \begin{equation}
     X_{\infty} = 2^9 \left(\sqrt{15} + \frac{98}{27} \right) 
     = \left(\frac{16 + 8 \sqrt{15} }{3} \right)^3.    
   \label{C.8}
   \end{equation}

The solutions $x_1 = x_1(l)$ give us unstable cosmological soutions (as $t \to \infty$), 
while  $x_2 = x_2(l)$ lead us to stable ones. 

Let us consider the second series of solutions.
Here, one can obtain more subtle relation instead of (\ref{C.5})
\begin{equation}
  x_2 = - \frac{6}{l} - \frac{3}{l^2} + o(l^{-2}),
 \label{C.5a}
 \end{equation}
as $l \to \infty$. This  relation implies the following
asymptotic formula   for the 
parameter of dimensionless variation of the  effective gravitational constant
in Jordan frame (see (\ref{r5.G0}))  
\begin{equation}
  \delta   =  \frac{3}{l} + o(l^{-1}),
\label{C.9}
\end{equation}
as $l \to \infty$. Thus, we get 
\begin{equation}
  \delta  = \delta(l) \to 0,
    \label{C.10}
\end{equation}
for $l \to \infty$.  The relation (\ref{C.10}) was discovered numerically 
in ref. \cite{IvKob}.  

\subsection{The solutions for $m=12$ and $l = 11$}

Let us consider the case $m=12$ and $l = 11$. We get
\begin{eqnarray}
     x(\nu) =  - \frac{1}{162} \bar{Y}^{1/2} - (55/54)  + \frac{\nu}{2} Z^{1/2},
          \label{D.0}   \\ 
    Z =  (5456/243) \bar{Y}^{-1/2} - X^{1/3}+ (299/6561)X^{-1/3}- (250/729),     \label{D.1}   \\
    \bar{Y} = 6561 X^{1/3}- 1125 - 299 X^{-1/3},  \label{D.2} \\
    X = (46/59049) \sqrt{1093} + 12673/531441,     \label{D.3}  
   \end{eqnarray}  
 where $\nu = \pm 1$. Approximate numerical values for 
 $x_1 = x(-1)$ and $x_2 = x(1)$ read
 \begin{equation}
   x_1= - 1.487006703,  \qquad  x_2 =  - 0.818209536.  \label{D.4}
 \end{equation}
The cosmological solution corresponding to $x_2$ is stable and gives 
the $\delta$-parameter (from (\ref{r5.G0})) 
 \begin{equation}
        \delta =   -3.049 \times 10^{-4}, \label{D.5}
 \end{equation}
 which obeys the bounds (\ref{r5.G1}). The solution corresponding to $x_2$ was found  numerically 
 in ref. \cite{IvKob}. 
 
 \subsection{The solutions for $m=11$ and $l = 16$}
 
  For $m=11$ and $l = 16$ we get two solutions. The first
 solution to the master equation, corresponging to unstable 
 cosmological solution,  reads  
  \begin{eqnarray}
   x_1 = X^{1/3} - (19967/509796) X^{-1/3} -  481/714,
   \label{D.6}  \\
   X = \sqrt{28457}/(49(34)^{3/2})- 5656195/36399434, 
   \label{D.7}    
 \end{eqnarray}
 or numerically,  $x_1 = - 0.871886679$. The second one
 was obtained in ref. \cite{IvKob}: 
 \begin{equation}
 x_2 = - 1/2.  
 \end{equation}
 It gives a zero variation of the effective gravitational constant $G$
 in Jordan frame, i.e. $\delta = 0$. The stability of the corresponding cosmological
 solution was proved earlier in \cite{ErIvKob-16}.

 \subsection{The solutions for $m=15$ and $l = 6$}
  
   Let us put $m=15$ and $l = 6$. We get two solutions. The first
   one corresponds to unstable   cosmological solution. It reads  
   \begin{eqnarray}
    x_1 = X^{1/3} - (2/9) X^{-1/3} - 8/3,
    \label{D.6}  \\
    X = \sqrt{187}/3^{3/2}-71/27, 
    \label{D.7}    
  \end{eqnarray}
  or numerically,  $x_1 = - 4.278163073$. The second one
  was obtained in ref. \cite{IvKob}: 
   \begin{equation}
  x_2 = - 2.  
  \end{equation}
  It leads to zero variation of $G$ ($\delta = 0$). The stability of 
  the corresponding cosmological  solution was proved in \cite{ErIvKob-16}.

\section{Conclusions}

We have considered the  $D$-dimensional  Einstein-Gauss-Bonnet (EGB) model
with  two non-zero constants $\alpha_1$ and $\alpha_2$.  
By using the  ansatz with diagonal  cosmological  metrics, we have studied 
 a class of solutions with  exponential 
time dependence of two scale factors, governed by two Hubble-like parameters $H >0$ and $h$, 
corresponding to submanifolds of dimensions $m > 2$ and $l > 2$, respectively, with  $D = 1 + m + l$. 
The equations of motion were reduced to the master equation 
$\lambda(x,m,l) = 0$ (see (\ref{r3.8R}) or (\ref{r3.13.Ma})), 
where the parameter $x = h/H$ obeys the restrictions:  $x \neq 1$,  $x \neq - m/l$ and $x \neq x_{\pm}$
($x_{-} < x_{+} < 0$) are defined in (\ref{r3.9D}).
By using our earlier analysis from ref. \cite{IvKob-18ml} we have proved 
that the master equation has  real solutions only for  $\alpha > 0$. In this case 
there are two solutions: $x_1$, $x_2$, which satisfy
 
 $$x_{-} <x_1 < - m/l < x_2 < x_{+} < 0.$$
  
The master equation may be solved in radicals, since it is equivalent to a polynomial equation of  
fourth order (for $l > 2$).  See Appendix.

Any cosmological solution corresponding to $x_1$ or $x_2$ (for  $\alpha > 0$)
 describes an exponential expansion of  3-dimensional subspace (``our'' space) with
the Hubble parameter $H > 0$ and anisotropic behaviour of $(m-3+ l)$-dimensional internal space:
expanding in $(m-3)$ dimensions (with Hubble parameter $H$) and   contracting 
in $l$ dimensions (with Hubble-like parameter $h$). 

By using our earlier results from ref. \cite{IvKob-18ml} we have proved that 
the solution corresponding to $x_2$ is stable  in a class of cosmological solutions with diagonal metrics, 
while the  solution corresponding to  $x_1$ is unstable. 

We have presented several 
examples of exact solutions (in terms of $x = h/H$) in the following cases: i)  $m =l$;  ii) $m=3$, $l =4$;
iii) $m = 9$, $l >2$; iv) $m = 12$, $l = 11$; v) $m=11$, $l=16$ and vi) $m =15$, $l =6$. 
 In case iii) we have also proved  the asymptotical relation for variation of $G$:
 $\dot{G}/(GH)   =  3/l + o(1/l)$,  as $l \to \infty$, which is valid for stable solutions.


\vspace{0.3truecm}
 
 {\bf Acknowledgments}

The publication has been prepared with the support of the "RUDN University Program 5-100" (recipient V.D.I., mathematical model development). The reported study was funded by RFBR, project number 19-02-00346 (recipient A.A.K., simulation model development). 


\renewcommand{\theequation}{\Alph{section}.\arabic{equation}}
\renewcommand{\thesection}{}
\setcounter{section}{0}

\section{Appendix}

The master equation (\ref{r3.13.Ma})  reads
\begin{equation}
F(x) = A x^4 + Bx^3 + Cx^2 + Dx + E = 0,
 \label{A.1}
\end{equation}
where
\begin{eqnarray}
 A =  (l-2)(l-1)l(l+1), \label{A.2A} \\ 
 B =  4(l-1)^2 l(m-1), \label{A.2B} \\ 
 C =  2(l-1)(m-1)(3lm-2m-2l), \label{A.2C} \\ 
 D =  4(l-1)(m-1)^2 m,   \label{A.2D} \\ 
 E =  (m-2)(m-1)m(m+1).   \label{A.2E} 
 \end{eqnarray}

By making the substitution
\begin{equation}
 X = u + d, \qquad  d = - B/(4A) = - (l-1)(m-1)/( (l-2)(l+1)),  \label{A.3}
\end{equation}
we get 
\begin{equation}
F (u + t) = (l-2)(l-1)l(l+1)(u^4 + a u^2 + b u + c),  \label{A.4}
\end{equation}
 where for our values of $A, B,C,D, E$ we have
\begin{eqnarray}
a = 2(m-1)(l^2 m-7 l m + 4m + l^3 - 4 l^2 + 7l)/( (l-2)^2 l(l+1)^2),
    \label{A.5a} \\  
b = (8(m-1)^2(l^2m-3lm+4m+l^3-4l^2+3l))/( (l-2)^3 l (l+1)^3), \
    \label{A.5b} \\  
c = -(m-1)(l^4 m^3 -5 l^3 m^3 + 19 l^2 m^3-39l m^3+32 m^3  \quad \nonumber \\ 
+ 2 l^5 m^2-13 l^4 m^2+37 l^3 m^2-79 l^2 m^2 + 101 l m^2 -56 m^2   \quad
\nonumber \\
+l^6 m-7 l^5 m+ 20 l^4 m - 48 l^3 m+107 l^2 m- 97 l m + 8m  \quad
\nonumber \\
+l^6-l^5-14l^4+38l^3-35l^2+11l)/ ((l-2)^4 (l-1) l (l+1)^4).  \quad  \label{A.5c}  
 \end{eqnarray}
 It may be readily verified that $b > 0$ for all $m > 2$ and $l > 2$.
 
 Then,  equation   (\ref{A.1}) reads as follows 
 \begin{equation}
 u^4 + a u^2 + b u + c = 0.  \label{A.6}
 \end{equation} 
 
 Solving the  equation   (\ref{A.6}) by the well-known Ferrari method needs an arbitrary (real) solution to
 the cubic equation 
  \begin{equation}
  y^4 + \frac{5}{2} a y^2 + ( 2 a^2 - c) u +  \frac{1}{2} (a^3 - a c - \frac{1}{4} b^2 )  = 0,  \label{A.7}
  \end{equation} 
 or another equation  
   \begin{equation}
    v^3 + p v +  q = 0,     \label{A.8}
    \end{equation}
 where  
 \begin{equation}
     y =  v  -  \frac{5}{6} a,     \label{A.9}
  \end{equation}
  and
 \begin{eqnarray}
  p =  -  \frac{1}{12} a^2 - c,
     \label{A.10p} \\  
  q =  -  \frac{1}{108} a^3 + \frac{1}{3} a c - \frac{1}{8} b^2.   
   \label{A.10q}  
  \end{eqnarray}
 
  It follows from relations (\ref{A.5a}), (\ref{A.5b}), (\ref{A.5c}) that
  \begin{eqnarray}
   p = (2(m-1)(m+l)(lm^2+2m^2+l^2m \nonumber \\
   -5lm-2m+2l^2-2l))/( 3(l-2)^2(l-1)l^2(l+1)^2), 
      \label{A.10pp} \\  
   q =(-4(m-1)^2 (m+l)(5lm^3+4m^3+10l^2m^2-42lm^2-4m^2 \nonumber \\
   +5l^3 m- 42l^2 m+73lm+4l^3-4l^2))/( 27(l-2)^3(l-1) l^3(l+1)^3).   
    \label{A.10qq}  
   \end{eqnarray}
  It may be readily verified  that $p > 0$ for all $m > 2$ and $l > 2$.
 
 The real  solution to cubic  equation (\ref{A.7}) 
   has  the following form  
   \begin{equation}
        y =  -  \frac{5}{6} a   - \frac{p}{3 U} +  U,      \label{A.11}
     \end{equation}
 where 
 \begin{equation}
       U  =  \sqrt[3]{ -  \frac{q}{2} + \sqrt{\frac{q^2}{4} + \frac{p^3}{27}}} .    \label{A.12}
 \end{equation}
 Here $U > 0$ since  $p > 0$.
 
 The  complex solutions to quartic equation (\ref{A.1}) read as follows
    \begin{equation}
   x = d + \varepsilon_1 \frac{1}{2} \sqrt{a + 2y} + \varepsilon_2 
   \frac{1}{2} \sqrt{- \left(3 a + 2y + \varepsilon_1 \frac{2b}{\sqrt{a + 2y}}\right)},      \label{A.13}
   \end{equation}
where $\varepsilon_1 = \pm 1$ and $\varepsilon_2 = \pm 1$ are two independent sign parameters.

Here $a + 2y > 0$, $b >0$  and we have two real roots which correspond to the following choice of sign 
\begin{equation}
  \varepsilon_1 = - 1.
         \label{A.14}
   \end{equation}

\end{document}